\documentclass{aa}  
\usepackage{xcolor}
\usepackage{graphicx}
\usepackage{txfonts}

\usepackage{amssymb}	
\providecommand{\abs}[1]{\lvert#1\rvert}

\usepackage{hyperref}
\usepackage{soul}
\usepackage{ulem}
\begin{document}

   \title{Dust-void evolution driven by turbulent dust flux can induce runaway migration of Earth-mass planets}
   \titlerunning{Dust-void evolution can induce runaway migration of Earth-mass planets}


   \author{R. O. Chametla
          \inst{1}
          ,
          O. Chrenko\inst{1}
          ,
          F. S. Masset\inst{2}
          ,
          G. D'Angelo\inst{3}
          \and 
          D. Nesvorn\'{y}\inst{4} 
          }
    \authorrunning{R. O. Chametla, et al.}
   \institute{Charles University, Faculty of Mathematics and Physics, Astronomical Institute.
V Hole\v{s}ovi\v{c}k\'ach 747/2, 180 00,
Prague 8, Czech Republic\\
              \email{raul@sirrah.troja.mff.cuni.cz}
         \and
             Instituto de Ciencias F\'isicas, Universidad Nacional Autonoma de M\'exico.
Av. Universidad s/n, 62210 Cuernavaca, Mor., M\'exico\\
        \and 
            Theoretical Division, Los Alamos National Laboratory, Los Alamos, NM 87545, USA\\
        \and 
            Department of Space Studies, Southwest Research Institute, 1050 Walnut St., Suite 300, Boulder, CO 80302, USA\\ 
        }

   \date{Received XXX; accepted YYY}

  \abstract
  {Torques from asymmetric dust structures (so-called dust-void and filamentary structures) formed around low-mass planets embedded in a non turbulent dust-gas disk can exceed the torques produced by the gas disk component, then governing the planet's orbital dynamics. Here, we investigate how these structures (hence the dust torque) change when the effect of turbulent dust diffusion and dust feedback are included, and the direct implications on the migration of Earth-like planets. Using the \textsc{Fargo3D} code, we perform 2D and 3D multifluid hydrodynamic simulations,
  focusing on a non-migrating planet with the mass $M_p=1.5\,M_\oplus$ in 2D and 
  on migrating planets with $M_p\in[1.5,12]\,M_\oplus$ in 3D.
  We vary the $\delta$-dimensionless diffusivity parameter in the range $[0,3\times10^{-3}]$ and consider three different Stokes numbers $\mathrm{St}=\{0.04,0.26,0.55\}$, which are representative of the gas, transitional and gravity-dominated regimes, respectively. In our 2D models, we find that turbulent diffusion of dust prevents the formation of the dust-void and filamentary structures when $\delta>3\times10^{-4}$. Otherwise, dust structures survive turbulent diffusion flow. However, dust and total torques become positive only in transitional and gravity-dominated regimes. In our 3D models, we find that the dust-void is drastically modified and
  the high-density ring-shaped barrier delineating the dust-void disappears if $\delta\gtrsim10^{-4} $, due to the effect of dust turbulent diffusion along with the back-reaction of the dust. For all values of $\delta$, the filament in front of the planet is replaced by a low-density trench. 
  Remarkably, as we allow the planets to migrate, the evolving dust-void can drive either runaway migration or outward (inward) oscillatory-torque migration. Our study thus suggests that low-mass Earth-like planets can undergo runaway migration in dusty disks.
  }


   \maketitle
%

\section{Introduction}
It is well established that protoplanetary disks are composed of gas and a small fraction of dust, with a mass of about $1\%$ of the total mass of the gas \citep[e.g.][]{Armitage2024}. However, a clear picture of the dynamics of these constituents has not yet emerged. For instance, there is still no general consensus on the possible mechanism that gives rise to the redistribution of angular momentum in protoplanetary disks. Since the last century, several efforts have been made to solve this problem. Currently, two mechanisms are considered as possible candidates to explain the transport of angular momentum in protoplanetary disks \citep[see][for a review]{Turner2014,Lesur_etal2022}: (i) Turbulence parameterized by a dimensionless $\alpha$-parameter \citep{SS1973}, which can be driven by different sources (for instance, magnetorotational instability \citep[MRI;][]{BH1991}, hydrodynamic instabilities such as the Rossby wave instability \citep[RWI;][]{Lovelaceetal1999,Lietal2000,Lietal2001} or the vertical shear instability \citep[VSI;][]{AU2004}. (ii) Magnetized disc winds \citep[MDW;][]{BP1982}.

In the frame of angular momentum transport driven by turbulence, the dynamics of the gas and dust in a protoplanetary disk are implicitly linked by the value of the $\alpha$-parameter, since a non-zero value of $\alpha$ generates a certain level of turbulence in the gas which in turn induces turbulent diffusion (i.e., stirring) of the dust through the action of drag forces \citep{YL2007}. Nevertheless, it is important to mention that dust coupling to gas turbulence can lead to diffusion \citep{CLin2020} and anti-diffusion \citep[i.e., dust clumping, see e.g.][]{J2007}. Both processes can affect the formation of planets \citep[e.g.][]{V2020}. Indeed, turbulent dust diffusion can prevent dust accumulation in regions of a protoplanetary disk where density or pressure gradients occur.

With all the above in mind, we are interested in studying the effects of turbulent dust diffusion on planetary migration. In recent 2D numerical studies, in which dust diffusion was not taken into account, it has been proposed that non-symmetric dusty structures may arise in the vicinity of a planet \citep[][]{BLlP2018,Regaly2020,Guilera2023,Chrenko2024}. The so-called dust-void and dust filament structures (which are regions of low and high dust density, respectively) can produce a total torque on the planet that, in many cases (depending on the Stokes number $\mathrm{St}$ of the dust and on the dust-to-gas mass ratio $\epsilon$), is positive and therefore leads to an outward migration for low-mass planets ($0.3\leq M_p\leq10M_\oplus$, with $M_\oplus$ the Earth mass). 

In this work, we examine the effect of turbulent dust diffusion on the density distribution of dust around a low-mass planet through multifluid simulations. In particular, we wish to analyse the formation of the dust-void and filament under the action of the turbulent diffusion of dust. We begin with 2D models, similar to those considered by \citet{BLlP2018}, to quantify the change in the torque exerted on the planet when dust can diffuse. Subsequently, we investigate the possible implications of 3D dust distributions on planetary migration. We do so by performing 3D multifluid simulations, parameterizing dust sedimentation through the $\delta$-diffusion coefficient \citep{MorfilV1984,Weber_etal2019}. We aim at determining whether the formation of a 3D dust-void and a filament structure are possible and whether gravitational torques are affected.

\begin{figure*}
 \centering  \includegraphics[width=1.0\textwidth]{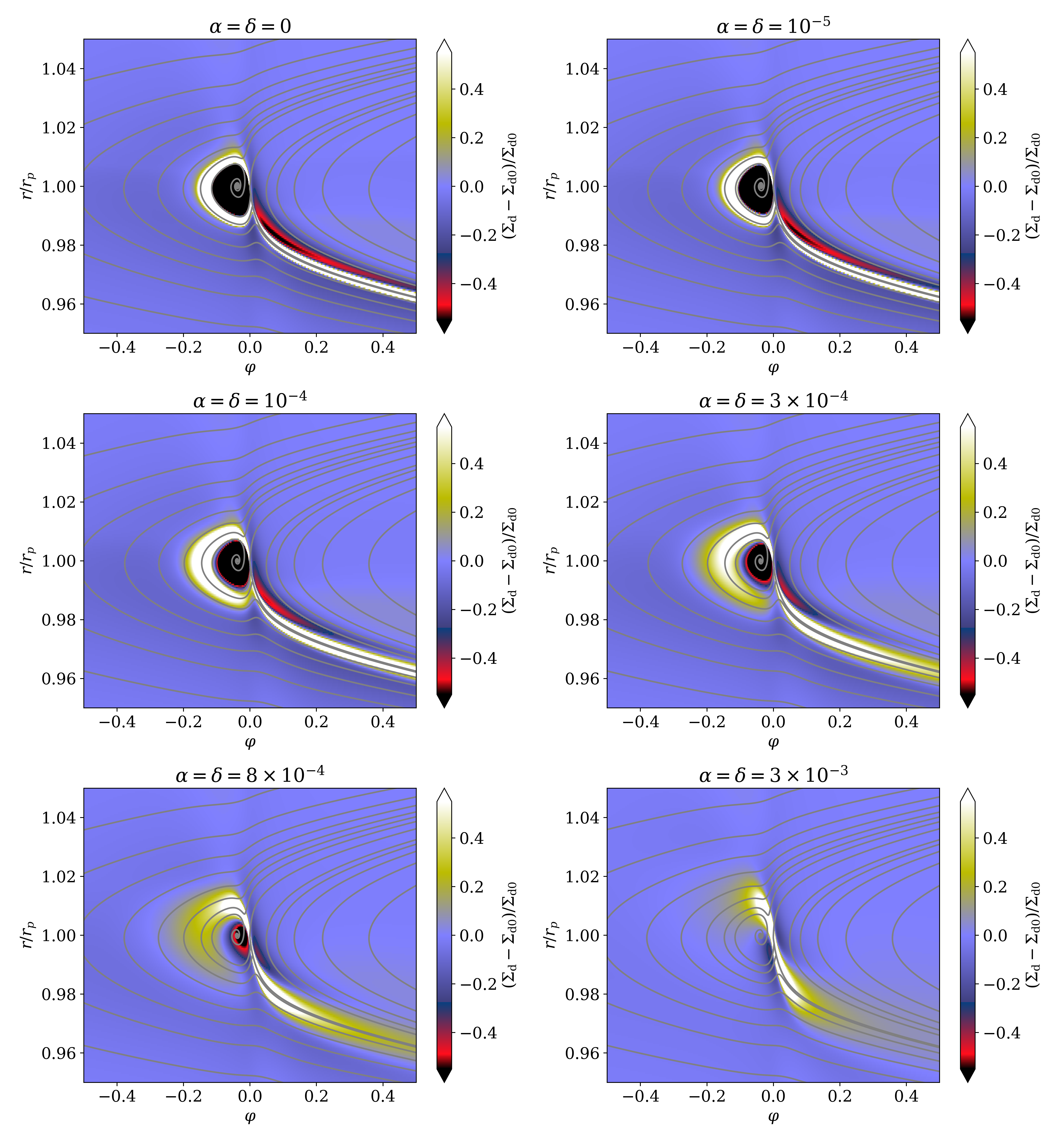}
  \caption{Relative perturbation of dust density for a Stokes number $\text{St}=0.55$ (which is representative of the gravity-dominated regime), generated by an Earth-mass planet ($M_p=1.5M_\oplus$). The solid gray lines show the dust streamlines. All images are shown for time $t=50$ orbits. Note that when the dust diffusion is not included (upper left panel), the formation of the dust-void behind the planet and the dust filament in front of the planet exhibit the same structure as reported in Fig. 1, case c), of Benítez-Llambay (2018)). However, when dust diffusion increases, the dust-void and filament eventually disappear (see bottom right panel). }
 \label{fig:dust_hole}
\end{figure*}

The paper is laid out as follows. In Section \ref{sec:model}, we describe the physical model of a dust-gas disk. The code and numerical setups used in our 2D and 3D multifluid simulations are provided in Section \ref{sec:setup}. In Section \ref{sec:dust_hole_mor}, we show the results of our 2D numerical models on dust-void destruction due to turbulent dust diffusion. The dust-void morphology obtained from 3D simulations and the effects on the migration of a low-mass planet are presented in Sections \ref{sec:3dRuns} and \ref{sec:runaway}. Finally, a brief discussion and our conclusions are given in Section \ref{sec:conclusions}.

\section{Physical Model}
\label{sec:model}
In this section, we describe different components of our physical model, including turbulent
dust diffusion and back-reaction of dust on gas employed in this study.
\subsection{Governing equations}
\label{sec:gas}
We consider a 3D, non-self-gravitating gas-dust disk whose evolution is governed by the following equations:
\begin{align}
\partial_t\rho_\mathrm{g}+\nabla\cdot(\rho_\mathrm{g} \mathbf{v})&=0,
\label{eq:gas_cont}\\
\partial_t(\rho_\mathrm{g}\mathbf{v})+\nabla\cdot(\rho_\mathrm{g}\mathbf{v}\otimes\mathbf{v}+p\mathsf{I})&=-\rho_\mathrm{g}\nabla\Phi+\nabla\cdot\mathrm{\tau}-\mathbf{f}_\mathrm{d},
 \label{eq:gas_mom}\\ 
\partial_t\rho_\mathrm{d}+\nabla\cdot(\rho_d \mathbf{u}+\mathbf{j}_\mathrm{d}
)&=0,
\label{eq:dust_cont}\\
\partial_t(\rho_\mathrm{d}\mathbf{u})+\nabla\cdot(\rho_\mathrm{d}\mathbf{u}\otimes\mathbf{u})&=-\rho_\mathrm{d}\nabla\Phi+\mathbf{f}_\mathrm{d},
\label{eq:dust_mom}
\end{align}
where $\rho_\mathrm{g}$, $\rho_\mathrm{d}$, $\mathbf{v}$, and $\mathbf{u}$ denote the gas density, the dust density, the gas velocity and the dust velocity, respectively. Furthermore, $\Phi$ denotes the gravitational
potential, $\mathsf{I}$ is the unit tensor, and $p$ is the gas pressure. For the latter, we consider the locally isothermal
equation of state
\begin{equation}
p=c_s^2\rho_\mathrm{g},
 \label{eq:pressure}
\end{equation}
where $c_s$ is the isothermal sound speed. Note that the sound speed is kept constant, i.e. we do not include an energy equation. Consequently, the aspect ratio of the gas disk $h_{\mathrm{g}}$ remains
constant. The aspect ratio of the dust $h_{\mathrm{d}}$ is set constant initially (Eq.~\ref{eq:h_dust}), at $t=0$.

To mimic the turbulence in the gaseous component, we include a viscous stress tensor in the right-hand side (RHS) of the Eq. (\ref{eq:gas_mom}), whose three dimensional form can be written as
\begin{equation}
\mathbf{\tau}=\rho_\mathrm{g}\nu\left[\nabla\mathbf{v}+(\nabla\mathbf{v})^{\mathrm{T}}-\frac{2}{3}(\nabla\cdot\mathbf{v})\mathsf{I}\right],
    \label{eq:visc}
\end{equation}
in which $\nu$ is the gas kinematic viscosity, parameterized through the usual \citet{SS1973} description
\begin{equation}
\nu=\alpha c_s H_\mathrm{g}.
    \label{eq:alph}
\end{equation}
In the equation above, $H_\mathrm{g}$ is the pressure scale height of the gas in the disk, defined as $H_\mathrm{g}=c_s/\Omega_\mathrm{Kep}$ in which $\Omega_\mathrm{Kep}$ is the Keplerian angular frequency. 
It is important to note that the $\alpha$-parameter, apart from playing a role in the dynamics of the gas, also affects the dynamics of the dust because it generates a constraint on dust diffusion, as we describe below.

\subsection{Dust diffusion flux}

To take into account the possible effect of gas turbulence on the dust dynamics, we include in the dust mass conservation equation (see Eq. \ref{eq:dust_cont}) the gradient in the dust concentration through the vector $\mathbf{j}_\mathrm{d}$ given as \citep[see][]{MorfilV1984}
\begin{equation}
\mathbf{j}_\mathrm{d}=-D_\mathrm{d}(\rho_\mathrm{g}+\rho_\mathrm{d})\nabla\left(\frac{\rho_\mathrm{d}}{\rho_\mathrm{g}+\rho_\mathrm{d}}\right).
    \label{eq:dd}
\end{equation}
The quantity $D_\mathrm{d}$ is the dust diffusion coefficient, which is related to gas turbulence through the Schmidt number
\begin{equation}
\mathrm{Sc}=\frac{\nu}{D_\mathrm{d}},
    \label{eq:Sch}
\end{equation}
which quantifies the relative effectiveness of the gas angular momentum transport and the dust-mixing processes. In this study we consider $\mathrm{Sc}=1$, since we analyze a dust fluid with a Stokes number below unity \citep[see][]{YL2007}. Note that, although both diffusive coefficients ($\nu$ and $D_\mathrm{d}$) have a common origin in the turbulence in the disk (and in this particular case they are even equal), they describe different physical processes. Therefore, in order to differentiate these two quantities, we will use the dimensionless diffusivity parameter
\begin{equation}
\delta=\frac{D_\mathrm{d}}{c_sH_\mathrm{g}},
    \label{eq:delta}
\end{equation}
which is analogous to the dimensionless $\alpha$-parameter.
\begin{figure*}
\centering
\includegraphics[width=1.0\textwidth]{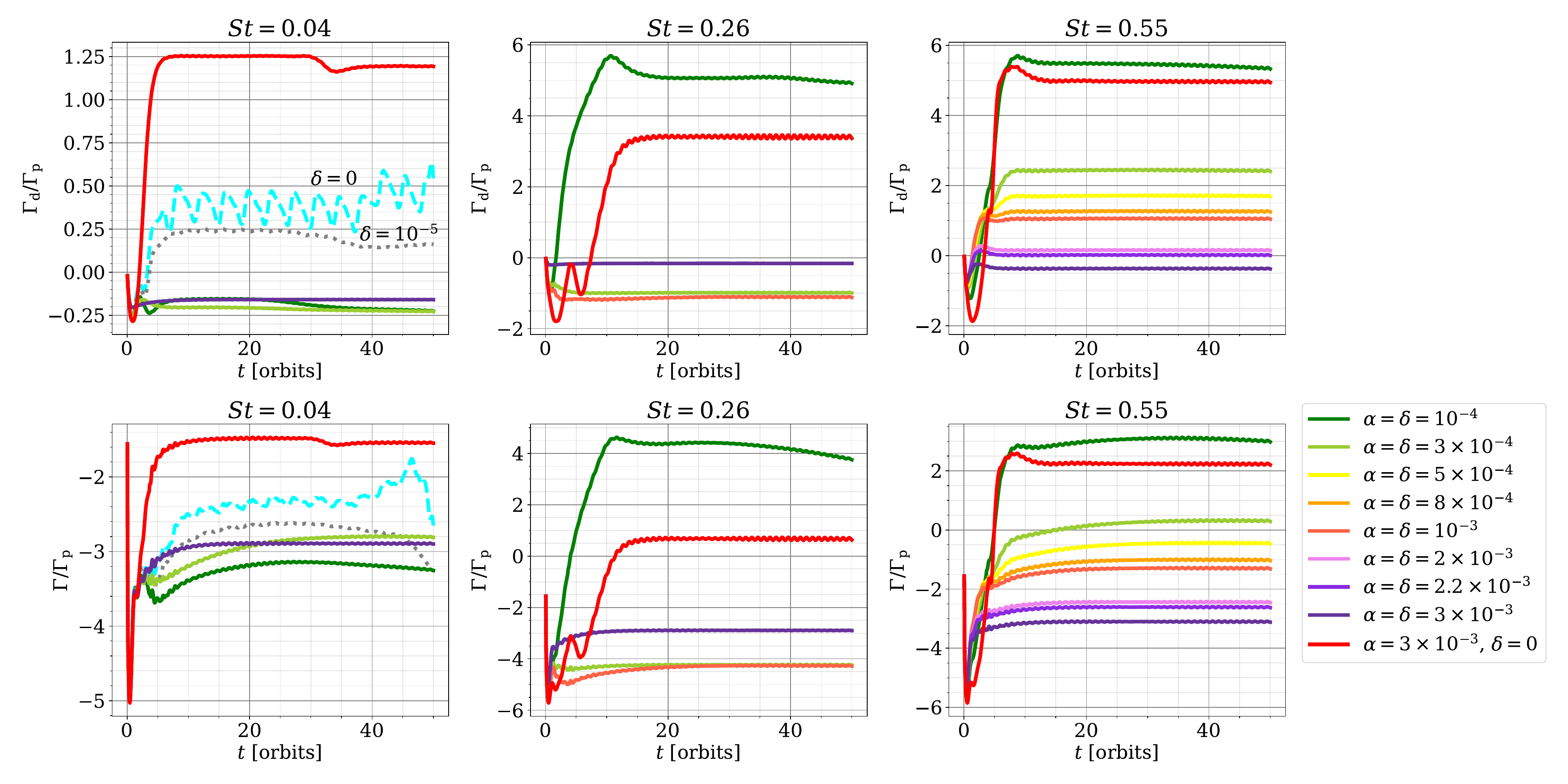}
 \caption{Temporal evolution of the dust (top row) and total (bottom row) torques on a planet of mass $M_p=1.5M_\oplus$ for representative cases of the gas (left), transitional (middle) and gravity-dominated (right) regimes.}
\label{fig:torques2d}
\end{figure*}
\subsection{Dust back-reaction force}

In Equations~(\ref{eq:gas_mom}) and (\ref{eq:dust_mom}), the last term on the RHS is the drag force term given as
\begin{equation}
    \mathbf{f}_\mathrm{d}=\frac{\Omega_\mathrm{Kep}}{\mathrm{St}}\rho_\mathrm{d}(\mathbf{v}-\mathbf{u}).
	\label{eq:dragforce}
\end{equation}
where $\mathrm{St}$ is a dimensionless measure of the dust–gas coupling that relates to the dust grain properties via
\begin{equation}   \mathrm{St}=\sqrt{\frac{\pi}{8}}\frac{r_\mathrm{m}\rho_\mathrm{m}\Omega}{\rho_\mathrm{g}c_s},
    \label{eq:St}
\end{equation}
where $r_{\mathrm{m}}$ is the grain size and $\rho_{\mathrm{m}}$ is the material density of the dust. Note that the effect of the feedback force term given in Eq. (\ref{eq:dragforce}) becomes more relevant in 3D disk models where dust settling leads to large concentrations of dust at the midplane of the disk.

\section{Multifluid hydrodynamical simulations}
\label{sec:setup}

To numerically solve Equations~\ref{eq:gas_cont}-\ref{eq:dust_mom}, we use the publicly available hydrodynamic
code \textsc{Fargo3D}\footnote{https://fargo3d.github.io/documentation}
\citep{BLlM2016,BLlKP2019} with an implementation of the
fast orbital advection algorithm
\citep[FARGO;][]{Masset2000} used in our 2D models, and with the new implementation of the rapid advection algorithm on arbitrary meshes \citep[RAM;][]{BLl2023} employed in our 3D simulations.

In both 2D and 3D multifluid simulations, the dust density profile is initialized from the gas density using the constant dust-to-gas mass ratio $\epsilon$ and composed of separate fluids with Stokes numbers of $\mathrm{St}=0.04,0.26$ and $\mathrm{St}=0.55$ at $r_0=1$. These Stokes numbers fall into the gas, transitional, and gravity-dominated regimes, respectively \citep[cases $a)$, $b)$ and $c)$ analyzed in][]{BLlP2018}. Although for small values of the Stokes number and gas viscosity (hence, small values of $\delta$, such as $10^{-4}$ used here) the feedback may be ineffective, when $\mathrm{St}$ approaches unity the feedback can produce significant changes in the radial velocity of the gas \citep{K2017}. Therefore, to get an idea of the effect of turbulent dust diffusion on the dust structures around of the planet, in case c) we do not include the feedback.

\subsection{Set-up of 2D simulations}
\label{subsec:2d-setup}
Our 2D numerical simulations are carried out following the framework of \citet{BLlP2018} (see their Section 2), which in turn is based on the gas-dust disk model of \citet{Webber_etal2018}, with the two major differences in the disk model mentioned above: we include the effect of the back-reaction of the dust on the gas and the diffusion of the dust. The effects of these two processes, neglected in the study of \citet{BLlP2018}, are presented in Section \ref{sec:dust_hole_mor}. Below we briefly recall some aspects of the initial conditions.

The 2D multifluid equations (that is, two-dimensional version of the set of Eqs. \ref{eq:gas_cont}-\ref{eq:dust_mom}) are solved
on a polar grid centred on to the star, and in a frame co-rotating with
the planet of orbital radius $r_p$. The surface density of the gas disk is chosen to be a power law of
  $r$:
\begin{equation}    \Sigma_g(r)=\Sigma_0\left(\frac{r}{r_p}\right)^{-\sigma},
    \label{eq:Sigma}
\end{equation}
with $\Sigma_0=2\times10^{-3}/\pi$ in code units (which corresponds to $\approx200\mathrm{g} \mathrm{cm}^{-2}$ in cgs units at $r_p=5.2$ au) and a surface density slope of $\sigma=1/2$. The temperature decays as a power law of the radius, as well, with slope $\beta=1$. We use $h_\mathrm{g}=0.05$ and we consider $\alpha\in[0,3\times10^{-3}]$. 
The dust density profile is initialized from $\Sigma_g$, using the constant dust-to-gas mass ratio $\epsilon=\Sigma_\mathrm{d}/\Sigma_\mathrm{g}=0.01$.

The gravitational potential $\Phi$ is composed of the gravitational potential of the star, the planet and the indirect term 
(arising from non-inertial forces, second term on the RHS in Eq. (\ref{eq:Planet_potential})). Additionally, in the gravitational potential of the planet, we apply a softening length $r_\mathrm{sm}$ equal to the planet Hill radius, $r_\mathrm{H}= r_p(M_p/3M_\star)^{1/3}$ (here $M_p$ and $M_\star$ are the planet and star masses, respectively), in order to avoid numerical issues. Note that the value of $r_\mathrm{sm}$, for the low-mass planets considered in this work, is significantly smaller than the value required in a gaseous disk to produce an approximate agreement between 2D and 3D torques \cite[e.g.][]{TO2024}, while it is comparable to or larger than the thickness of the dusty disk. This entails that we overestimate the gaseous torque and underestimate the dusty torque, so that our statements about the importance of the dusty torque are conservative. We applied the same cut-off of the torque inside the Hill sphere and we also considered the same smoothing length as in \citet{BLlP2018} for comparison purposes, since our objective is to describe the effects of diffusion and feedback on dust distribution. However, we are aware that said dust distribution is sensitive to the choice of the smoothing length in 2D \citep[][]{Regaly2020,Guilera2023}, whose value is not well-determined \citep[for a detailed discussion see][]{Chrenko2024,HY2024}.

Unless otherwise specified, the planet has a mass $M_p=1.5M_\oplus$ and is held on a fixed circular orbit around a star of the mass $M_\star=1M_{\odot}$. We model the entire ring of a disk ($2\pi$ in azimuth) with a radial extent between $r=0.48$ and $r=2.08$, considering $N_r=3130$ radial and $N_\phi=12288$ radial and azimuthal uniform grid zones, respectively.

\subsection{3D disk simulations}
\label{subsec:3d-setup}
The setup of our 3D model is based on the gaseous disk of \citet[][see their Appendix A]{MB2016}.
We use spherical coordinates $(r,\theta,\phi)$, where $r$ is the radial distance from the star, $\theta$ is the polar angle ($\theta=\frac{\pi}{2}$ at the midplane of the disk), and $\phi$ is the azimuthal angle.
The aspect ratio of the gas disk is $h_\mathrm{g} = H_\mathrm{g}/r$. The gas disk density is given by
\begin{equation}
\rho_\mathrm{g}=\rho^\mathrm{eq}_\mathrm{g}(\sin\theta)^{-\beta-\xi+h_\mathrm{g}^{-2}}
    \label{eq:rhog}
\end{equation}
with 
\begin{equation}
\rho^\mathrm{eq}_\mathrm{g}=\frac{\Sigma_0}{\sqrt{2\pi}h_\mathrm{g}r_p}\left(\frac{r}{r_p}\right)^{-\xi},
    \label{eq:rhoeq}
\end{equation}
where $\Sigma_0$ is the surface density at $r=r_p$. We set $\beta=1$, $\xi=1.5$ and $h_\mathrm{g}=0.05$ in all models considered here. We initialize the gas velocity components as follows: $v_r=v_\theta=0$ and 
\begin{equation}
v_\phi=\sqrt{\frac{GM_\star}{r\sin{\theta}}-\xi c_s^2.}
 \label{eq:vphi}
\end{equation}
The gravitational potential $\Phi$ is given by
\begin{equation}
\Phi=\Phi_S+\Phi_p,
 \label{eq:potential}
\end{equation}
where
\begin{equation}
\Phi_S=-\frac{GM_\star}{r},
 \label{eq:Star_potential}
\end{equation}
and
\begin{equation}
\Phi_p=-\frac{GM_p}{\sqrt{r'^2+r_\mathrm{sm}^2}}+\frac{GM_pr\cos{(\phi-\phi_p)}\sin{\theta}}{r_p^2}
 \label{eq:Planet_potential}
\end{equation}
are the stellar and planetary potentials, respectively. In
Eq. (\ref{eq:Planet_potential}), $r'=\abs{\mathbf{r}-\mathbf{r}_p}$ is
the cell-planet distance, with the planet located at $(r_p,\theta_p,\phi_p)=(1,\frac{\pi}{2},0)$ and, $r_\mathrm{sm}$ is softening length of the planet's potential. The second
term on the right-hand side of Eq. (\ref{eq:Planet_potential}) is the
indirect term arising from the reflex motion of the star. Our simulations were
performed with $r_\mathrm{sm} = 0.05H_\mathrm{g}$, which is comparable to twice the cell size of our numerical grid 
(we have done other experiments with a larger $r_\mathrm{sm}$ and found similar results). In our 3D models, no cut-off of the torque is applied inside the Hill sphere.

Since dust grains above the midplane are not pressure-supported, the vertical gravitational
force drives their settling on a timescale which is much shorter than 
that of viscous gas evolution. Consequently, the scale heights of dust and gas
can be different ($h_\mathrm{d}\neq h_\mathrm{g}$).
Therefore, we consider an initial dust density profile that depends on the dust pressure scale $h_\mathrm{d}$, such that
\begin{equation}
\rho_\mathrm{d}=\rho^\mathrm{eq}_\mathrm{d}(\sin\theta)^{-\beta-\xi+h_\mathrm{d}^{-2}}
    \label{eq:rhod}
\end{equation}
and 
\begin{equation}
\rho^\mathrm{eq}_\mathrm{d}=\epsilon\frac{\Sigma_0}{\sqrt{2\pi}h_\mathrm{d}r_0}\left(\frac{r}{r_p}\right)^{-\xi}.
    \label{eq:rhodeq}
\end{equation}
In Eq. (\ref{eq:rhodeq}), the initial dust-to-gas mass ratio $\epsilon$ is such that $\Sigma_\mathrm{d}/\Sigma_\mathrm{g}=0.01$. The dust pressure scale is defined through the dimensionless diffusivity parameter $\delta$ as
\begin{equation}
h_\mathrm{d}= h_\mathrm{g}\sqrt{\frac{\delta}{\delta+\mathrm{St}}}.
    \label{eq:h_dust}
\end{equation}

The radial and vertical components of the dust velocity are both zero initially, 
whereas the azimuthal component is set to start from a Keplerian rotation profile. Contrary to the 2D models, the planet is allowed to migrate in all of our 3D simulations.
In order to include the aerodynamic back-reaction (feedback) from dust on gas (cases with $\mathrm{St}=0.04$ and $\mathrm{St}=0.26$), we performed simulations using two fluids at a time (gas and one dust species).

\subsubsection{Computational grid}
\label{subsubsec:grid}

Due to the computational cost of 3D multifluid simulations, in order to achieve a numerical resolution similar to that used in the 2D models (see section \ref{subsec:2d-setup}), we use the \textsc{Fargo3D} code with the RAM module. This module allows us to use a non uniform mesh, with smaller cells near the planet \citep{BLl2023}. Such approach helps increase the resolution around the dust-depleted region at a more affordable computational cost. The numerical grid covers the disk region $[r_\mathrm{min},r_\mathrm{max}]=[0.48r_p,2.08r_p]$ in the radial directions,
$[\theta_\mathrm{min},\theta_\mathrm{max}]=[\frac{\pi}{2}-h_p,\frac{\pi}{2}]$ in co-latitude (we only simulate one hemisphere of the disk and assume symmetry in the other). The azimuthal extent is
$[\phi_\mathrm{min},\phi_\mathrm{max}]=[-\pi,\pi]$, covering the whole extent of the disk. 

Under this numerical scheme, the bumps in the radial and azimuthal mesh functions \citep[see Appendix B in][]{BLl2023} are parameterized by considering the constants $a=0.2618$, $b=0.3141$ to model the resolution transitions, and with $c=1.5$ and $c=15$ for the radial and azimuthal resolution contrast between the two levels, respectively. Vertical resolution and spacing
are chosen based on the dust aspect ratio (see Eq. \ref{eq:h_dust}). When $h_\mathrm{d}\geq0.1h_\mathrm{g}$, we use a uniform grid with constant resolution in the polar direction. Setting the number of grid cells $(N_r,N_{\theta},N_{\phi})=(1864,100,1864)$
leads to cube-sized cells 
$\Delta r=r_p\Delta \theta=r_p\Delta \phi=5\times 10^{-4}$ at the planet position. We also note that the grid attains a resolution of $H/100$. Otherwise, if $h_\mathrm{d}<0.1h_\mathrm{g}$, we use a mesh density function in the polar direction whose resolution ratio is specified in Appendix \ref{sec:appendix}.

\subsubsection{Boundary conditions}
\label{subsubsec:boundaries}

We use boundary conditions for each dust species similar to those implemented for the density and velocity components of the gas \citep[see][for details]{ChM2021}. We use damping boundary zones as in \citet{dVal2006},
the width of the inner and outer damping rings being $0.15r_p$ and $0.5r_p$, respectively. The damping time-scale 
at the edge of each ring is equal to $0.3$ of the local orbital period. We emphasize that the radial extension of the domain is large enough to avoid any disturbances from the disk boundaries affecting the dust void
region. Any perturbations found in this region are therefore purely a result of the dynamic interaction
between the gas, dust, and the planet. Additionally, we use damping zones only in the radial direction
because their inclusion in the vertical direction would drastically modify dust settling 
and might lead to an overly large dust-to-gas mass ratio in the midplane.
Lastly, since we consider only one hemisphere of the disc, we use reflecting
boundary conditions at the midplane.

\begin{figure}
\includegraphics[width=0.4853\textwidth]{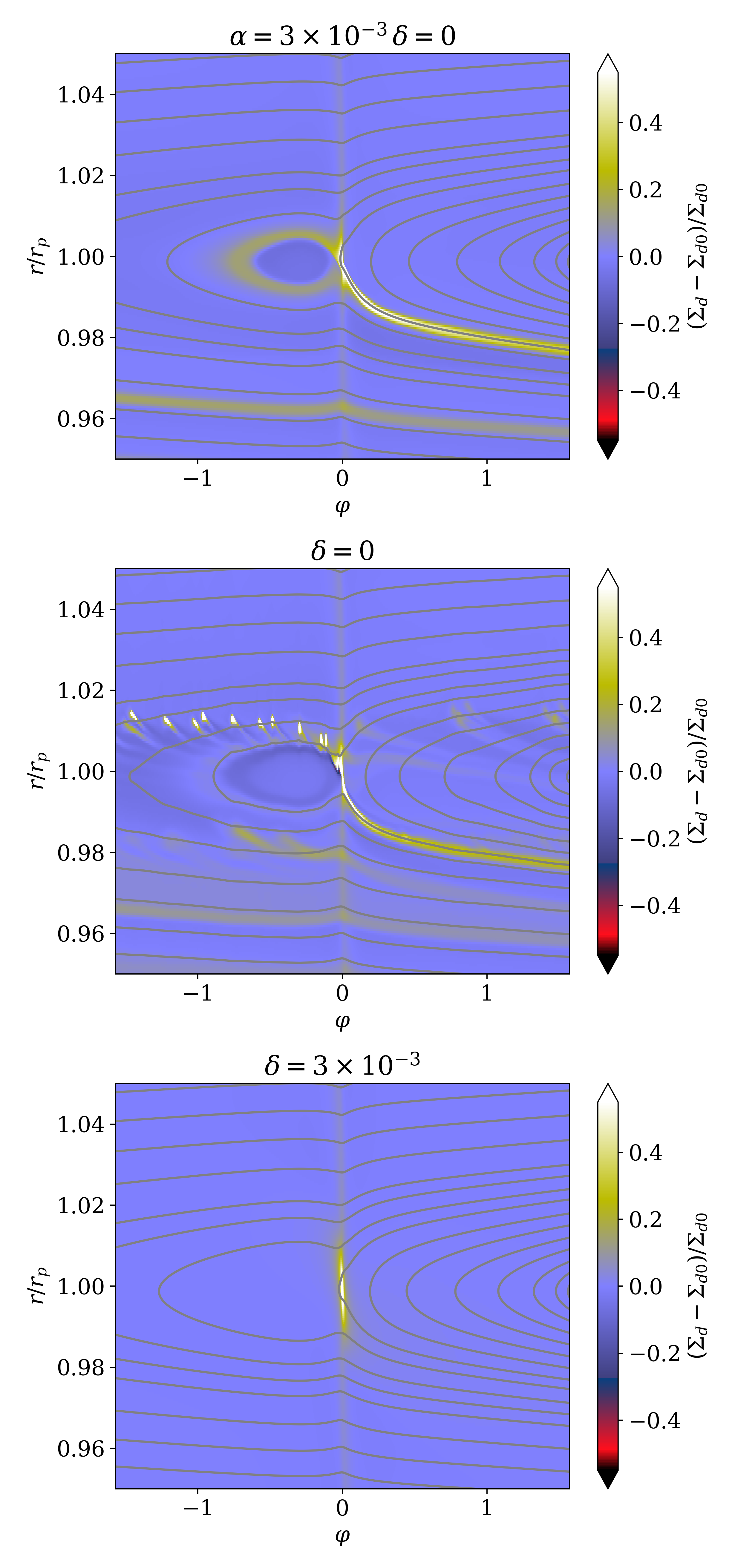}
 \caption{Two-dimensional dust-void structure for three different values of $\delta$ in the gas-dominated regime ($\mathrm{St}=0.04$), for a planet with $M_p=1.5M_\oplus$. \textit{Top.} Case without dust turbulent diffusion in a viscous gas disk, which is similar to the case a) studied in \citet[][their figure 1]{BLlP2018}. \textit{Middle.} Case of an inviscid gaseous disk ($\alpha=0$) without dust turbulent diffusion ($\delta=0$). \textit{Bottom}. Dust-void and filament are destroyed for $\delta=3\times10^{-3}$. Snapshots at  $t=50$ orbits.}
\label{fig:St04}
\end{figure}

\begin{figure}
\includegraphics[width=0.45\textwidth]{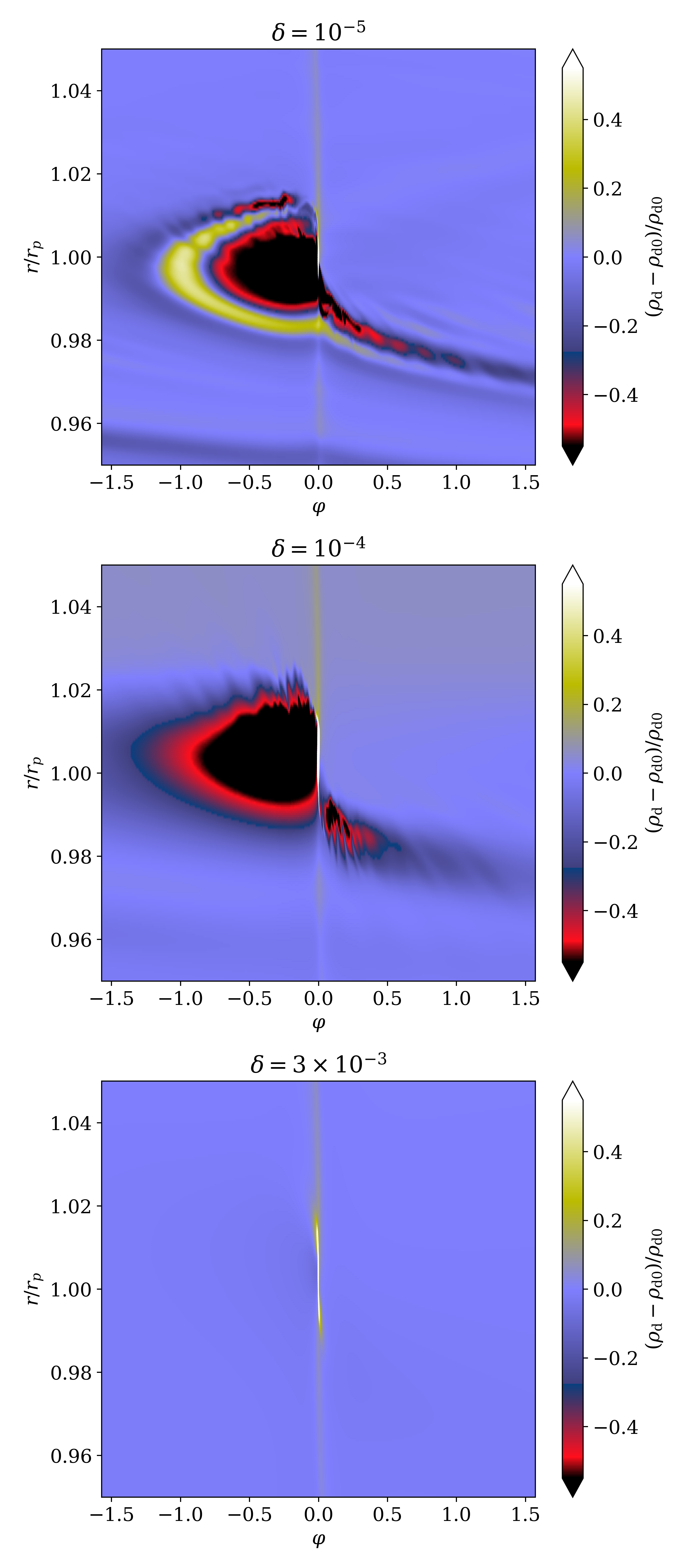}
 \caption{Three-dimensional dust-void structure for three different values of $\delta$ in the gas-dominated regime ($\mathrm{St}=0.04$), for a planet with $M_p=1.5M_\oplus$. \textit{Top.} Snapshot (calculated at $t=70$ orbits) in the case of $\delta=10^{-5}$. A 
 low-density trench forms in front of the planet instead of a filament.
 \textit{Middle.} Case with $\delta=10^{-4}$. The dust-void structure resembles the shape found in 2D simulations but without the high-density ring-shaped barrier. \textit{Bottom}. Dust-void and filament disappear for $\delta=3\times10^{-3}$ (snapshot calculated at $t=100$ orbits).}
\label{fig:St043D}
\end{figure}


\section{Destruction of the 2D dust-void in a gaseous turbulent disk}
\label{sec:dust_hole_mor}

In this section, we describe how the 2D dust density structures near
a $M_p=1.5M_\oplus$ planet change when the effects of turbulent diffusion and dust feedback are included. 
For the purpose of maintaining a
clearer notation throughout this study, we
omit stating the value of $\alpha$ when $\delta$ is specified because these parameters are identical.

\subsection{Effect of turbulent dust diffusion}
\label{subsec:2dturbulentdif}

To gain insight into the effect of turbulent dust diffusion, we first present the results of the models with $\mathrm{St}=0.55$, in which the dust feedback on the gas is neglected and the influence of the turbulent flow diffusion can be isolated more clearly.

In Fig. (\ref{fig:dust_hole}) we show the relative perturbation of dust density and the dust streamlines around the planet in the case of a Stokes number $\mathrm{St}=0.55$ and six different values of $\delta$, calculated at $t=50$ orbits. When $\delta=0$, a dust-void structure appears behind the planet, this feature is encircled by a high-density ring-shaped region, while in front of the planet a filamentary structure is formed. Looking at the other panels, we find that the dust-void structure and the dust filament undergo minor modifications for values of $\delta< 3\times10^{-4}$. 

As viscosity and turbulent dust diffusion increase, the distribution of dust around the planet is altered.
When $\delta\geq3\times10^{-4}$, the size of the dust-void reduces and the filament in front of the planet starts to vanish. At a value of the diffusivity parameter $\delta=3\times10^{-3}$, we find that neither the dust-void structure nor the filament form. This last case deserves particular attention since \citet[][]{BLlP2018} reported formation of the dust void and filament for the gas viscosity $\alpha=3\times10^{-3}$ (and $\delta=0$; see their figure 1).

This result is noteworthy because the value of $\alpha$ used by \citet{BLlP2018} is actually comparable with recent estimates of turbulence levels inferred from observed disks \citep{Flaherty_2020,Jiang_2024}. Therefore, turbulence dust diffusion can be critically important in these situations. It should be emphasized that we recover the same structure as case c) reported in \citet{BLlP2018} when we consider $\alpha=\delta=0$, that is, in a dusty gaseous disk without turbulence (see top left panel of Fig. \ref{fig:dust_hole}).

The trend appearing in Fig. (\ref{fig:dust_hole}) can be understood as a competition between two antagonic processes. The distribution of the dust results from its advection (an inward drift and a horseshoe U-turn) and diffusion due to its interaction with the turbulent gas. The dust settles into a pattern whose characteristics depend on the timescale for diffusion (which tends to fill the dust void) and that of advection (which tends to result in a dust-void). When the timescale for the former is short enough, i.e. when $\delta$ is sufficiently large, the dust-void is suppressed.

\begin{figure*}
\centering
\includegraphics[width=1\textwidth]{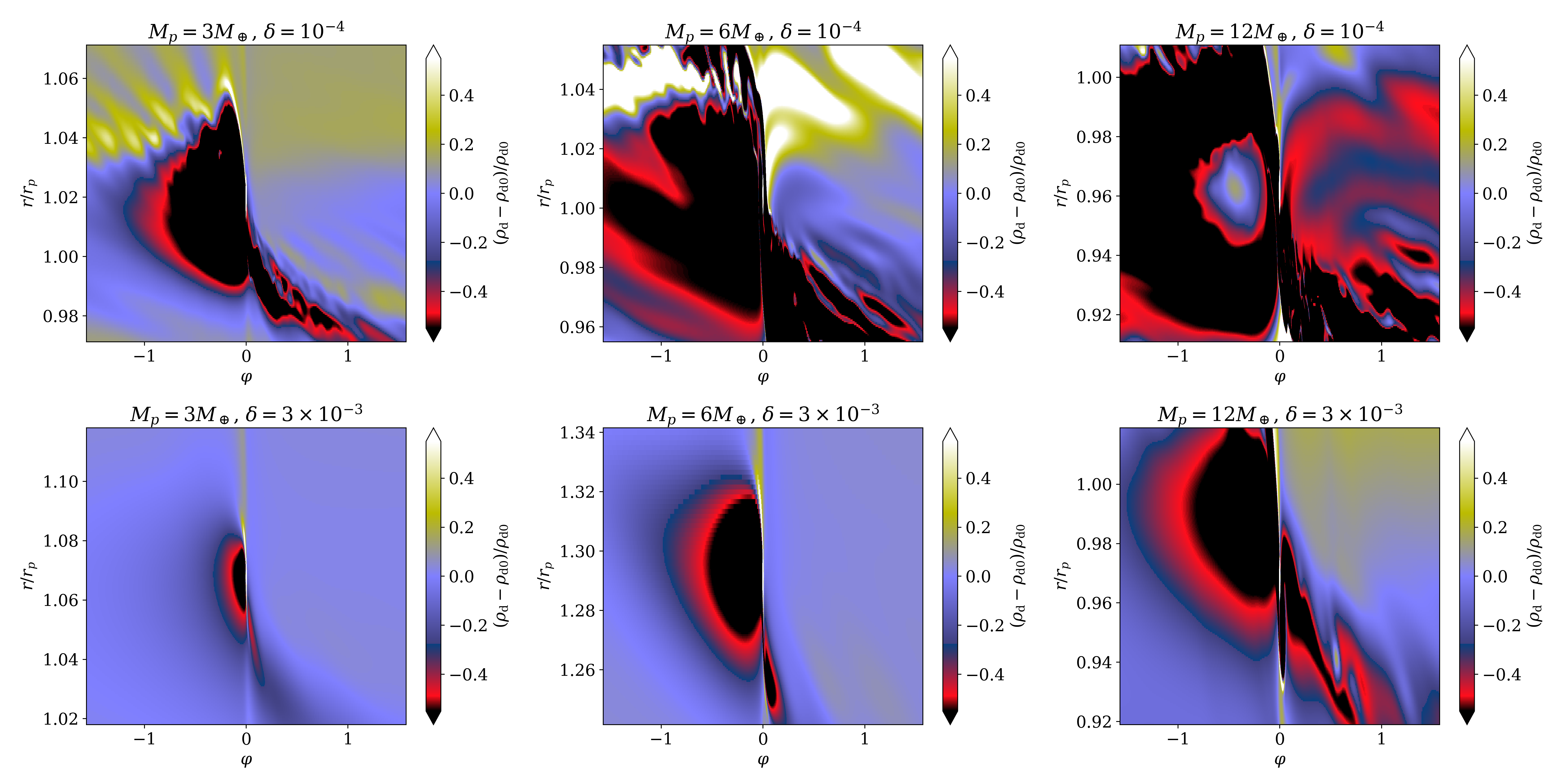}
 \caption{Relative perturbation of dust density at the midplane of the disk for $\alpha=\delta=10^{-4}$ (top) and $\alpha=\delta=3\times10^{-3}$ (bottom) calculated at $t=100$ orbits. Three different planetary masses in the transitional regime ($\mathrm{St}=0.26$) are considered (see the labels of individual panels). Note, the planet is located at the center of each plot.}
\label{fig:lMp}
\end{figure*}

\subsection{Lengthening and weakening of high-density dust-void barrier: dust back-reaction force effect}
\label{subsec:2dfeedback}

Similarly, for a Stokes number $\mathrm{St}=0.26$ (transitional regime not shown here), we find again that the dust-void and the filament structure begin to disappear for $\delta\geq3\times10^{-4}$ and are absent when $\delta\simeq3\times10^{-3}$. We emphasize that when the dust-void and filament structures are maintained, which happens for $\delta<3\times10^{-4}$, their shape and extension resemble those found in the gravity-dominated regime. In other words, the radius of the ring-shaped high-density structure that delimits the dust-void increases as the diffusivity parameter decreases. Dust from the surroundings cannot penetrate this region because of aerodynamic friction by the gas and scattering by the planet. Additionally, we find that the high-density filament in front of the planet becomes more prominent as the parameter $\delta$ reduces.
\begin{figure}
\includegraphics[width=0.4\textwidth]{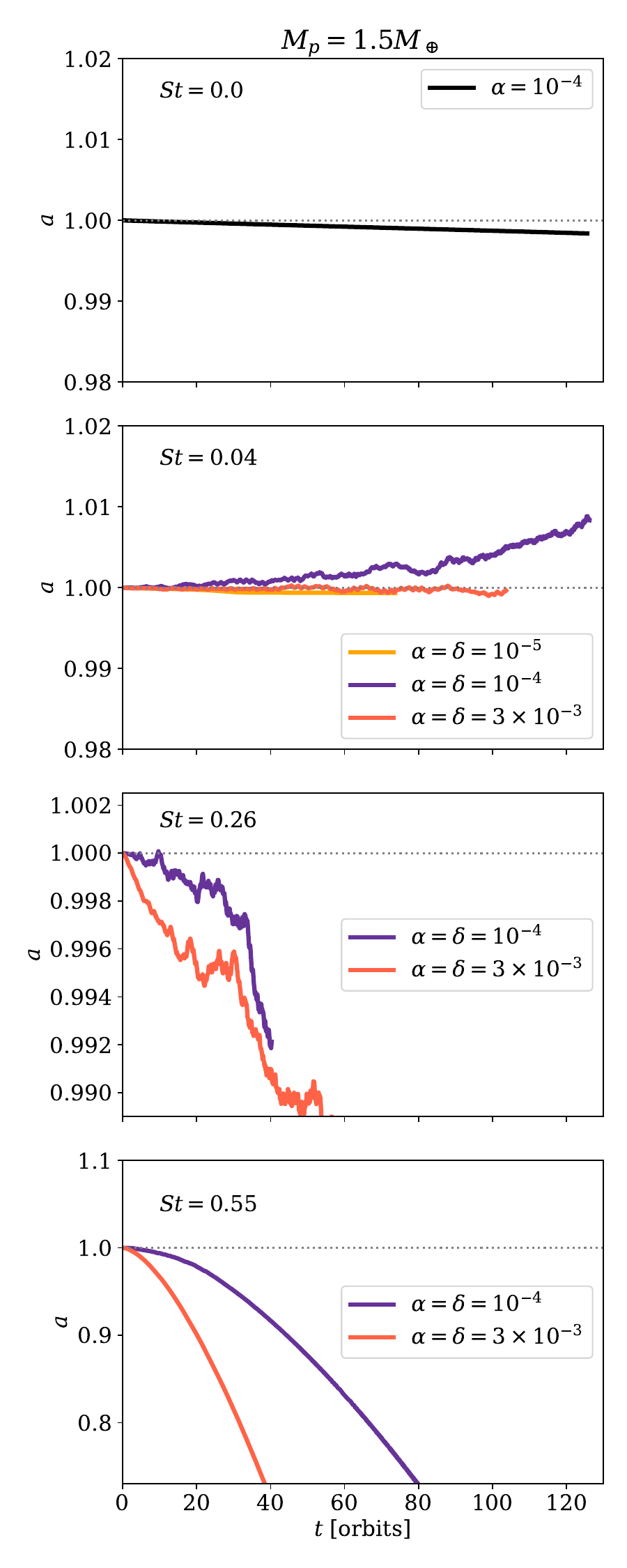}
 \caption{Temporal evolution of the semimajor axis of a planet with $M_p=1.5M_\oplus$ embedded in a purely gaseous 3D disk (top) and in 3D dusty disks representative of the 
 gas, transitional, and gravity-dominated regimes ((second, third, and bottom panel, respectively).}
\label{fig:semi}
\end{figure}
\begin{figure}
\includegraphics[width=0.5\textwidth]{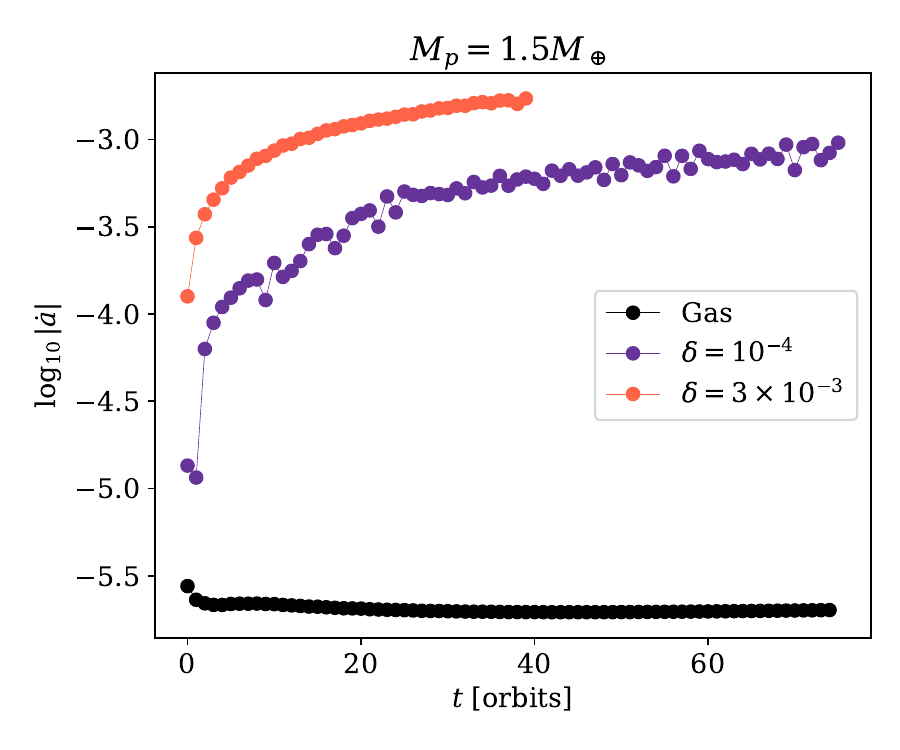}
 \caption{Temporal evolution of the migration rate $\dot{a}$ for a migrating planet of $M_p=1.5M_\oplus$ embedded in a gaseous disk, and also in a dusty-gaseous disk with $\delta=10^{-4}$ and $\delta=3\times10^{-3}$ for gravity-dominated regime ($\mathrm{St}=0.55$) without feedback of the dust on the gas.
 }
\label{fig:opMp1_5}
\end{figure}

In general, the explanation for the widening of the dust-void and the elongation of the filamentary structure lies in the fact that, the feedback and the turbulent diffusion of the dust incidentally conspire to counteract the scattered dust flow generated by the planet. That is, turbulent diffusion produces a spreading of the ring-shaped structure, but due to the dispersion produced by the planet the dust tends only to diffuse away from the planet. However, because the dust exerts a drag force on the gas, the pressure of the gas increases in the region where the dust moved, generating a new pressure maximum that prevents the dust located at the outer edge of the ring from escaping. This then results in a ring-shaped structure with a larger radius. A similar effect occurs in the filament in front of the planet, while turbulent diffusion tries to spread the dust, the scattering generated by the planet forces the dust to move away from it, but the feedback counteracts it, leading to an elongation of said structure.

For the case when the dust is strongly coupled to the gas (case $\mathrm{St}=0.04$, i.e., gas-dominated regime), we find that the dust-void and filament structures are drastically modified compared to the case when turbulent dust diffusion and dust feedback are discarded (see first and second panels in Fig. \ref{fig:St04}). For the case $\delta=0$, the dust barrier delimiting the dust-void is dragged towards the edges of the gas horseshoe region, where there may be vortices propagating because the gas disk is inviscid (since $\alpha=0$). A similar effect occurs in the dust filament formed in front of the planet, since some of the dust approaching the filament from upstream is carried toward the edge of the horseshoe region (see second panel in Fig. \ref{fig:St04}). Therefore, the dust-void is left unprotected which leads to a shallower dust-void and the dust filament is weakened. 

Lastly, if there is a non-zero turbulent diffusion of the dust, the blurring of the dust-void and the filament is stronger and they are completely erased when $\delta=3\times10^{-3}$ (see lower panel in Fig. \ref{fig:St04}).

\subsection{Impact of the dust-void on the planet's torque}

An immediate consequence of the dust-void and filamentary structures not surviving turbulent dust diffusion is reflected in the torque that the dust exerts on the planet. In the case of gravity-dominated regime ($\mathrm{St}=0.55$, without feedback), unlike \citet{BLlP2018}, we find that the dust torque on the planet is negative when $\alpha=3\times10^{-3}$. Note that the total torque is also negative (see the third column in Fig. \ref{fig:torques2d}). For a lower level of gas turbulence (which implies a lower value of $\delta$), the dust torque becomes less negative and it is possible to obtain a null total torque when $\delta\simeq3\times10^{-4}$. Reducing the diffusivity further, we find that for $\delta<3\times10^{-4}$ the total torque on the planet is always positive. This last result is relevant in regimes of low turbulence in the gas of protoplanetary disks \citep[e.g.,][]{Pinte_2016,Lesur_etal2022,Jiang_2024} as it entails that planetary migration can be directed outwards.

In the case of the transitional regime ($\mathrm{St}=0.26$, with feedback included), due to the broadening and lengthening of the dust-void and filament, respectively, we find similar values of the dust torque as in the gravity-dominated regime. For instance, see the cases with $\delta=10^{-4}$ and $\delta=3\times10^{-3}$ of the second and third columns in Fig. \ref{fig:torques2d}. However, it is worth mentioning that in the interval $3\times10^{-4}\leq\delta\leq10^{-3}$ the dust torque is negative for $\mathrm{St}=0.26$, which is different from the behavior of the torque in the gravity-dominated regime over the same range of $\delta$ values (where the dust torque is positive and its magnitude increases when $\delta\to0$, see Fig. \ref{fig:torques2d}). 

Finally, it is important to note that when the dust-void barrier is broken or destroyed, as it occurs in the case of the gas-dominated regime ($\mathrm{St}=0.04$), the dust torque is positive only for very small values of the $\delta$-parameter ($\sim$$10^{-5}$). Therefore, even in the presence of fine dust, the total torque on the planet is negative.

\section{3D dust-void morphology}
\label{sec:3dRuns}

We now focus on the results of the 3D multifluid simulations. Here, in addition to the case of $M_p=1.5M_\oplus$ considered in Sect.~\ref{sec:dust_hole_mor},
we extend our analysis towards more massive planets. Additionally, we
let the planets migrate instead of keeping them on fixed orbits.

\subsection{Dust-void destruction}

Fig. \ref{fig:St043D} shows the relative perturbation of dust density around a planet of $M_p=1.5M_\oplus$ at the midplane of the disk for the gas-dominated regime ($\mathrm{St}=0.04$). We have chosen three different values of the dust turbulent diffusion $\delta=10^{-5}$, $10^{-4}$ and $3\times10^{-3}$ as references cases that reveal the main differences in the dust density distribution
compared to 2D models.

When $\delta=10^{-5}$, the dust-void starts to show a shape resembling its 2D counterpart. However, the ring-shaped high-density structure at the edge of the dust-void is spread out, mainly due to the effect of turbulent dust diffusion. Note that, even though the feedback effect is less efficient in this case because the dust is initially confined to a very thin layer ($h_\mathrm{d}\ll h_\mathrm{g}$) there is a sufficient contribution from this to maintain dust accumulation leading to a process similar to that explained in the subsection \ref{subsec:2dfeedback}. Also, the ring-shaped high-density structure exhibits perturbations characteristic of an epicyclic motion when $r>r_p$ \citep[see][]{MN2012}. Additionally, we find that the filament in front of the planet is erased and a trench emerges in its place.

In the case $\delta=10^{-4}$, the dust void forms again but, unlike the 2D dust-void (see Figs. \ref{fig:dust_hole} and \ref{fig:St04}), it exhibits low central density and
is no longer delimited by a ring-shaped barrier of high density. Additionally, the trench and the epicyclic oscillations (in both structures) arise as for $\delta=10^{-5}$.
Note that the edges of the dust-void and trench are wider compared to those of the ring-shaped barrier and filament regions of the 2D models. This is likely a consequence of turbulent dust diffusion, which spreads out dust density gradients.

Lastly, when $\delta=3\times10^{-3}$, the size of the dust-void and the trench are considerably reduced, which means that both structures are filled by "fresh" dust. Seen from another perspective, dust scattering is only efficient very close to the planet, resulting in a negligible dust-void and shallower trench. The absence of the dust-void and filamentary 
structures for $\delta=3\times10^{-3}$ is similar in both 2D and 3D models (see bottom panels in Figs \ref{fig:St04} and \ref{fig:St043D}).

\subsection{Deformation and pollution of the dust-void for planets with a larger mass}

Now, we analyze the evolution of dust structures in 3D disks when the mass of the planet increases. For this proposal, we run a set of simulations with $M_p=3,6$ and $12M_\oplus$. We set the Stokes number to $\mathrm{St}=0.26$.

In Fig. \ref{fig:lMp}, we show the results of these numerical experiments. The relative perturbation of dust density at the midplane for a planet with $M_p=3M_\oplus$ exhibits a well defined structure for the dust-void and the filament, for both values of dust diffusion coefficients. However, a detailed inspection of the $\delta=10^{-4}$ case shows that the filament becomes a trench that connects directly to the dust-void. Another interesting feature is the overdense stripes adjacent to the dust-void and the trench in front of it. This dust pattern may be a consequence of buoyancy resonances (Chametla et al. in preparation) since an isothermal dusty gaseous disk behaves like an adiabatic pure gaseous disk \citep{LY2017}, in which the buoyancy resonances produce a similar stripe-like pattern in gas \citep[see][and references therein]{Ziampras2023}.

For a planet of $M_p=6M_\oplus$, when $\delta=10^{-4}$, the trench in front of the planet opens up enough to cover a considerable region compared to the dust-void, which in turn becomes polluted with material orbiting at $r>r_p$. Otherwise, the dust-void and the trench structures are well defined when $\delta=3\times10^{-3}$. That is, there is no material entering the dust-void region and the width of the trench does not increase as in the previous, lower-mass case. Lastly, for the most massive planet considered in this study (i.e., $M_p=12M_\oplus$) and a value of $\delta=10^{-4}$, we find again that the trench in front of the planet widens and it also contains some filaments of dust within it. Interestingly, an overdense island forms within the dust-void which survives until the end of the simulation. However, at a larger turbulent dust diffusion $\delta=3\times10^{-3}$, the dust-void remains clear and the trench is divided from the dust-void by a very thin filament that emerges from the planet position.

\begin{figure*}
\includegraphics[width=1\textwidth]{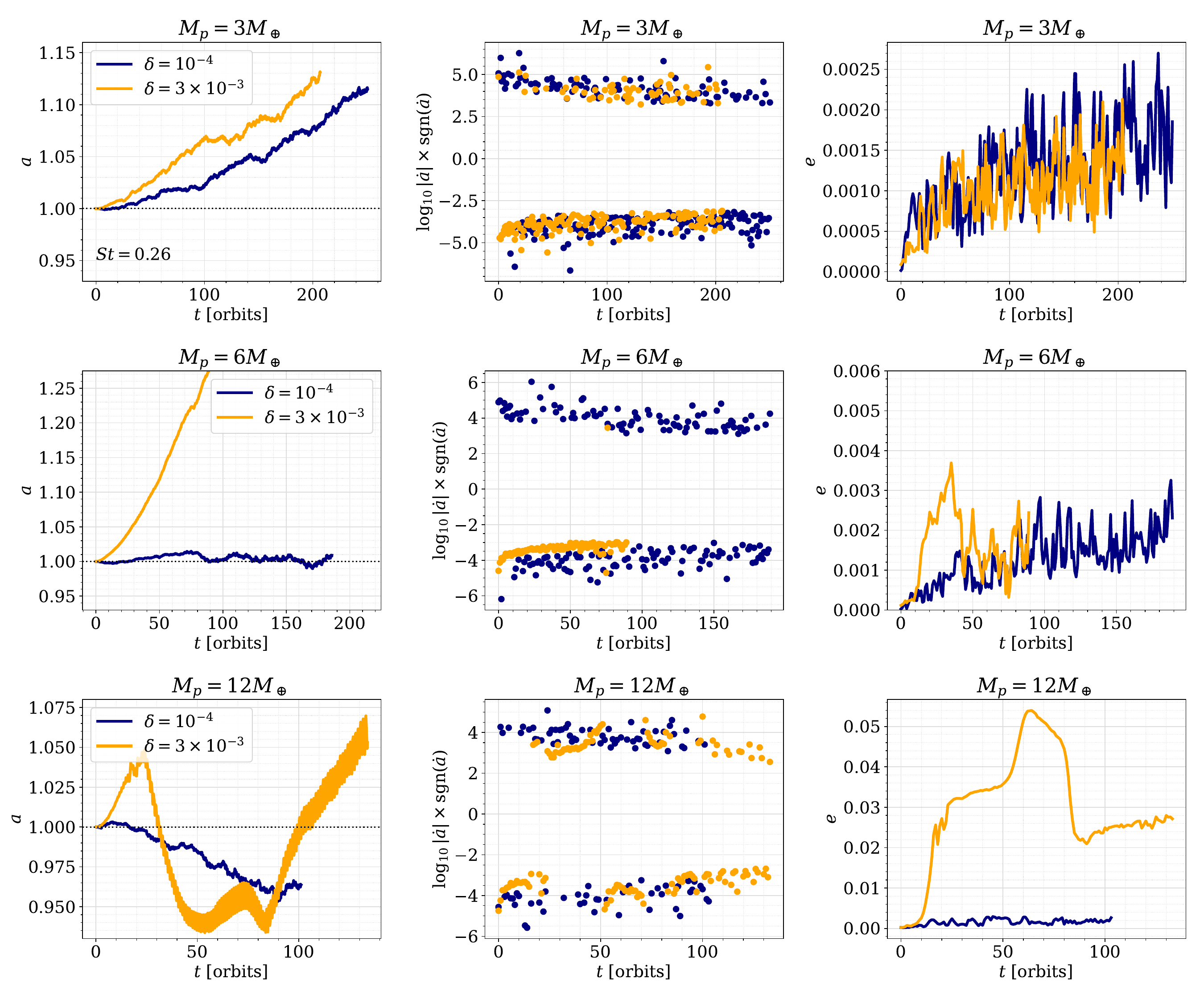}
 \caption{ Temporal evolution of the orbital parameters for the planets of larger mass migrating in a $3D$ dusty disk including $\delta=10^{-4}$ and $\delta=3\times10^{-3}$ for the transitional regime with $\mathrm{St}=0.26$. \textit{Left}. Semimajor axis $a$. \textit{Center}. Migration rate $\dot{a}$. \textit{Right}. Planet's eccentricity $e$.}
\label{fig:adte}
\end{figure*}

\section{Runaway and oscillatory migration driven by dust-void evolution}
\label{sec:runaway}

\subsection{Runaway Migration Signature}

Runaway migration of planets in a gas disk (without dust component) is a self-sustaining mechanism with short migration time-scales $\leq O(10^2)$ orbits \citep{Masset2003}, and it applies mainly to super-Earth, sub-Jovian and Jupiter-type planets \citep{Masset2003,P2008a,Pb2008,Pc2008,LinP2010,Paar14}. The main engine of the gas-driven runaway migration is a nonlinear corotation torque maintained by the change in the fluid elements' orbital radius executing U-turns at the end of horseshoe orbits. This torque scales with the rate of migration of the planet and with the vorticity-weighted coorbital mass deficit $\delta m$ \citep{Masset2003}. In the large mass regime, that deficit essentially corresponds to a coorbital mass deficit, which is proportional to the difference between the surface density trapped in the coorbital region and that of the orbit-crossing flow, while in the low mass regime, it essentially amounts to a jump of vortensity between the horseshoe region and the surrounding disk \citep{Paar14,McNally2018}.

In the first regime, two necessary conditions for runaway migration to occur in a gaseous disk are that the planet creates a partial gap and that the mass deficit $\delta m\approx M_p$. However, note the masses of the planets ($M_p\in[1.5,12]M_\oplus$) considered in this study are not large enough to satisfy both conditions simultaneously. Therefore, the migration of these low-mass planets in a gas disk must fall within the type I migration regime \citep[see][and references therein]{Ch2023}. As a proof of this last argument, in the upper panel of Fig. \ref{fig:semi} we show the temporal evolution of the semi-major axis of a planet with $M_p=1.5M_\oplus$ obtained from a 3D dust-free model. Clearly, the planet's semi-major axis $a$ does not change in a runaway fashion, but it rather varies at a constant rate over the first $150$ orbital periods as occurs in the type I migration regime over small orbital displacements.

The situation can be much different when a planet migrates in a dusty gaseous disk. Let us first analyze the migration for the case of $M_p=1.5M_\oplus$ and $\mathrm{St}=0.55$. In the bottom panel of Fig. \ref{fig:semi}, it can be seen that the semi-major axis of the planet changes considerably in a time less than a hundred orbital periods. The latter suggests that the planet is experiencing a runaway migration. To confirm this hypothesis, Fig. \ref{fig:opMp1_5} shows the temporal evolution of the planet's drift rate when the planet is embedded in a gas and gaseous-dusty disk with $\delta=10^{-4}$ and $\delta=3\times10^{-3}$, respectively. Clearly, the drift rate of the planet migrating in a gaseous-dusty disk is higher and varies considerably over the first orbital periods, even if the semi-major axis itself does not vary significantly during that short time frame. This suggests that the drift rate is not a function of the semimajor axis only, but also of the drift rate itself, as occurs in runaway migration. Note that the drift rate in a gaseous disk is constant which is to be expected due to the behavior of the planet's semi-major axis (see upper panel in Fig. \ref{fig:semi}). 

\subsection{Planetary drift rate as oscillatory-torque migration trace in dusty gaseous disks}

Now we return to analyze the migration in the cases when $\mathrm{St}=0.04$ and $\mathrm{St}=0.26$ shown in the second and third panels of Fig. \ref{fig:semi} for a planet of $M_p=1.5M_\oplus$. Except for the case of $\delta=3\times10^{-3}$ in the gas-dominated regime ($\mathrm{St}=0.04$), which exhibits a almost stagnant oscillating migration, we found that in these models the behavior of the semimajor axis is far from type I migration and in turn does not change as quickly while maintaining the same direction as in the case of runaway migration found in the gravity-dominated regime. 

A detailed inspection of each of the semimajor axes shows a migration with several changes of direction in short time compared to the dynamical timescale, which resembles stochastic migration \citep{NP2004,Nelson2005,BL2010}. However, it is important to emphasize that there is no global turbulence on the dust distribution. The changes in migration are mainly due to deflection and deformation of the dust-void and the trench, respectively, producing an oscillatory behavior in the torque felt by the planet.

Note that due to the time over which the 3D-multifluid simulations are run, a convergence study of the running time-averaged torque is beyond the scope of this work. Therefore, we henceforth prefer to adopt the term \textit{oscillatory-torque migration} to refer to such changes in planetary migration rather than stochastic migration.

Remarkably, for planets with a larger mass ($M_p\in\{3,6,12\}M_\oplus$) in the transitional-dominated regime ($\mathrm{St}=0.26$) also we find an oscillatory-torque migration. Fig. \ref{fig:adte} shows the temporal evolution of the semimajor axis, the migration rate and the eccentricity for these models. In the first column of Fig. \ref{fig:adte} it can be seen that the temporal evolution of the planet's semi-axis exhibits frequent oscillations within an orbital period for both values of dust turbulent diffusion. Although there seems to be a marked trend with respect to the mass of the planet, since for $\delta=10^{-4}$ as the mass of the planet increases, migration goes from outward to inward migration. On the other hand, for $\delta=3\times10^{-3}$ even when the mass of the planet increases the migration is directed outwards.

As a proof of oscillatory behavior in planet migration, in the second column of Fig. \ref{fig:adte} we show the drift rate of these planets with a larger mass. In these plots we have included the sign of the drift rate, since if there are mixed positive and negative values over the dynamical time it means that the planet experiences episodes of oscillatory migration. In all cases we find positive and negative values in the drift rate of the planet. However, particular attention must be paid to the case of $M_p=6M_\oplus$ and $\delta=3\times10^{-3}$, we find that for $t\leq70$ orbital periods the semimajor axis grows rapidly in a runaway fashion. We also observe that the sign of the drift rate does not change on tims time interval. This therefore suggests that for this particular case the planet first experiences a runaway migration and subsequently an oscillatory-torque migration.

Lastly, in the third column of Fig. \ref{fig:adte} we show the temporal evolution of the eccentricity. It can be seen that although there is an oscillatory behavior, there is also a marked increasing tendency of the eccentricity in almost all models. Except in the case when $M_p=12M_\oplus$ and $\delta=10^{-4}$ where there is an inward oscillatory-torque migration. Interestingly, the above suggests that an outward oscillatory-torque migration inevitably results in the development of planetary eccentricity.

\subsection{Triggering oscillatory-torque migration}

As we saw above, a planet migrating in a disk of gas and dust can experience either a runaway migration or an oscillatory-torque migration. Both types of migration are governed by the level of turbulent diffusion and feedback of the dust which shapes the structure of the dust-void and the trench. For instance, in the gas-dominated regime ($\mathrm{St}=0.04$) when $\delta=10^{-5}$, the dust-void is partially surrounded by broken ring-shaped high-density structure and the trench shows an elongated pattern (see Fig. \ref{fig:St043D}). These features lead to an inward oscillatory-torque migration.
On the other hand, for $\delta=10^{-4}$ we find that the dust-void and the trench feature survive, driving an outward oscillatory-torque migration. Finally, for $\delta=3\times10^{-3}$ the dust-void and the trench are significantly weakened and the planet experiences a nearly stagnant oscillatory-torque migration.

Furthermore, in the case of $\mathrm{St}=0.55$ (gravity-dominated regime) where the back-reaction force of the dust on the gas is not included, we find that the dust-void and the trench maintain a well-defined shape, resulting in an inward runaway migration for both values of $\delta=10^{-4}$ and $\delta=3\times10^{-3}$, respectively. Therefore, it appears from the models described above that the outcome of the runaway migration is mainly governed by the evolution of the dust-void. 
In fact, unless the dust-void is deformed or populated with dust islands, a runaway migration occurs. Otherwise, any modification of the dust-void inevitably
results in an oscillatory-torque migration.

\section{Discussion and Conclusions}
\label{sec:conclusions}

In this work, we report the results of 2D and 3D multifluid hydrodynamics simulations
including dust turbulent diffusion, aerodynamic dust feedback, and planet migration (in the 3D case). Our aim was to determine how the inclusion of aforementioned effects changes
the so-called dust-void and filament dust structures formed behind and in front of low-mass planets.

The importance of analyzing the possible effects of turbulent dust diffusion and feedback on these structures in the dust lies in the fact that, recently \citet{BLlP2018} have shown that the asymmetric distribution of dust around low-mass planets embedded in 2D disks can produce a torque that dominates the other components of the gas torques (corotation and Lindblad torques). In a subsequent study, \citet{Guilera2023} quantifies the impact due to the distribution of dust on the migration of the planets. However, in both studies they consider 2D-laminar disks, discarding the effect of turbulent diffusion of dust and also the dust feedback. It is worth mentioning that, in both studies the viscosity of the gas parameterized by the alpha coefficient was included, which was set to $\alpha=3\times10^{-3}$.

With this in mind, we have started our study with 2D models for a low-mass planet ($M_p=1.5M_\oplus$) on fixed circular orbit embedded in a dusty-gaseous disk following the same framework as in \citet{BLlP2018}. Nevertheless, here we have taken into account the turbulent dust diffusion (parameterized by $\delta\in[0,3\times10^{-3}]$)  and the back-reaction force from the dust on the gas. We find that, the dust-void and the filament are formed when $\delta\lesssim10^{-4}$. On contrary, higher levels of
turbulent dust diffusion prevent the formation of structures around the planet, reducing
the ability of dust-driven torques to cause outward migration.

Of course, there could be other factors that modify the 2D dust-void apart from the turbulent diffusion and feedback of dust. For instance, the smoothing radius of the planet's potential \citep{Regaly2020} or the absence of pebble accretion \citep{Chrenko2024}. The first, beyond being a physical variable, arises as a limitation of 2D numerical models since it is chosen ad hoc. Therefore, in order to obtain a more appropriate physical description of the evolution of the dust structures a 3D numerical treatment is necessary.

Consequently, we have included in our study 3D simulations of migrating planets with $M_p\in[1.5,12]M_\oplus$ (which also includes the representative case analyzed in the 2D models) which reveal that, the morphology of the dust-void and the filament depends strongly on the turbulent dust diffusion and also on the inclusion of feedback. The changes in the dust structures can be summarized as follows:
\begin{itemize}
\item[$\blacksquare$] In the case of smaller values of dust turbulent diffusion, $\delta<10^{-4}$, the dust-void is drastically modified and the ring-shaped high-density structure surrounding it is weakened and scattered. Note that, these structures still do not resemble the structures found at \citet{BLlP2018}, where neither turbulent dust diffusion nor feedback were included.
\item[$\blacksquare$] For a $\delta$ value of the order of $\sim10^{-4}$ the dust-void survives. 
It will simply depend on how much dust falls into the dust-void. 
\item[$\blacksquare$] In the case of $\delta$ values larger than $10^{-4}$ the dust-void is almost filled for lowest mass planet and, for larger masses, the dust- void decreases considerably.
\item[$\blacksquare$] Instead, for all values of $\delta$, the high-density filament in front of the planet is replaced by a trench.
\end{itemize}

Remarkably, we find runaway migration or an outward (inward) oscillatory-torque migration driven by the dust-void evolution due to the gas feedback and turbulent dust diffusion. It must be emphasized that, the evolution of the dust-void and trench depends on the $\delta$ value, the dust feedback and slightly on the mass of the planet. Lastly, the direction of the runaway migration (inward/outward) depends on the evolution of the trench, which is intrinsically linked to the turbulent diffusion and feedback of the dust.

\begin{acknowledgements}
     This work was supported by the Czech Science Foundation
    (grant 21-23067M). The work of O.C. was supported by the Charles University Research Centre program (No. UNCE/24/SCI/005). Computational resources were available thanks to the Ministry of Education, Youth and Sports of the Czech Republic through the e-INFRA CZ (ID:90254).
\end{acknowledgements}

\begin{appendix} 

\section{Vertical dust height resolution treatment}
\label{sec:appendix}

From Eq. (\ref{eq:h_dust}) it follows that the vertical height of the dust depends on the Stokes number and the coefficient of turbulent diffusion of the dust. Therefore, we have that for the pair of values of $(\mathrm{St}, h_\mathrm{d})=(0.04,3\times10^{-3})$ and $(\mathrm{St}, h_\mathrm{d})=(0.55,10^{-4})$ studied in our dusty gaseous disks, the aspect ratio of the dust takes the respective maximum and minimum values $h_\mathrm{d}^\mathrm{max}=0.013$ and $h_\mathrm{d}^\mathrm{min}=6.7\times10^{-4}$. This last value of $h_\mathrm{d}$ is of the order of the resolution value with cube-sized cells $\Delta r=r_p\Delta \theta=r_p\Delta \phi=5\times 10^{-4}$ at the planet position when a uniform mesh is used in the vertical component.

\begin{figure}
    \centering
    \includegraphics[width=1.0\linewidth]{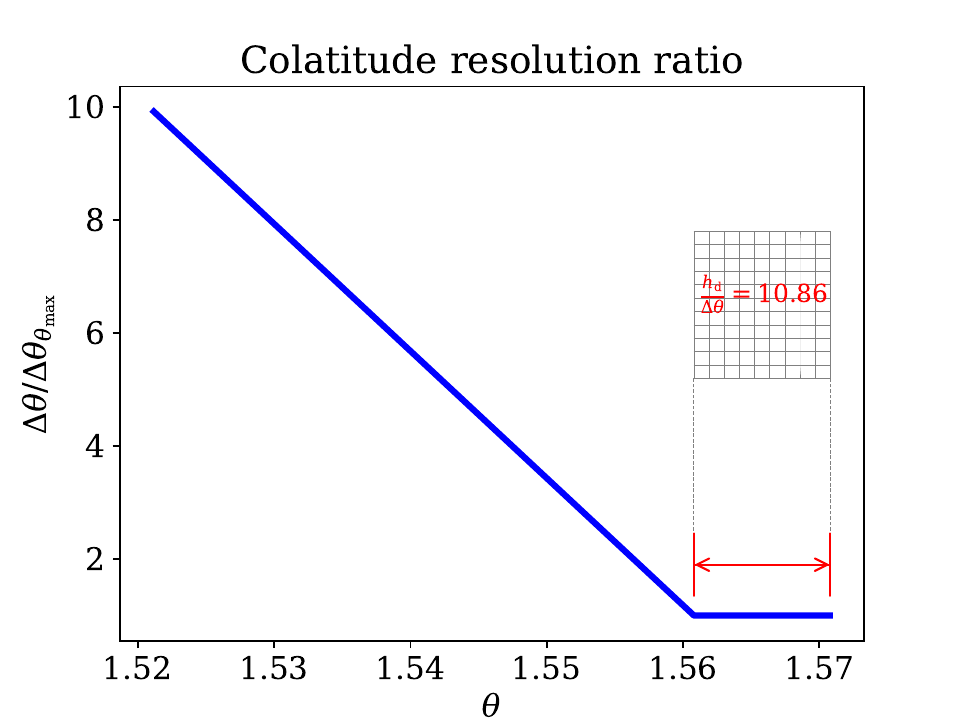}
    \caption{Colatitude resolution ratio for the models with a dust aspect ratio less than $0.1h_\mathrm{g}$. Note that, in the domain between $\theta=\frac{\pi}{2}-0.01$ and $\theta=\frac{\pi}{2}$ the resolution is constant and it satisfies the minimum requirement $h_\mathrm{d}/\Delta\theta=10.86$ in each model.}
    \label{fig:ratio}
\end{figure}

To avoid possible numerical effects on the dust dynamics around the planet, for the resolution of vertical component of the computational mesh we have adopted two prescriptions: 1) a uniform vertical domain with a constant resolution $\Delta\theta=5\times10^{-4}$ and, 2) A non-uniform vertical domain where the resolution increases considerably for values of $h_\mathrm{d}<0.1h_\mathrm{g}$ such that, it is satisfied that $h_\mathrm{d}/\Delta\theta\geq10.68$. In Fig. \ref{fig:ratio} we show the resolution ratio at colatitude used in 2). 
Within the domain of the computational mesh between $\frac{\pi}{2}-0.01$ and $\frac{\pi}{2}$, the resolution is kept constant. To maintain the aforementioned constraint, the number of zones $N_\theta$ is modified according to the value of $h_\mathrm{d}$ in each model.

\begin{figure}
    \centering
    \includegraphics[width=1.0\linewidth]{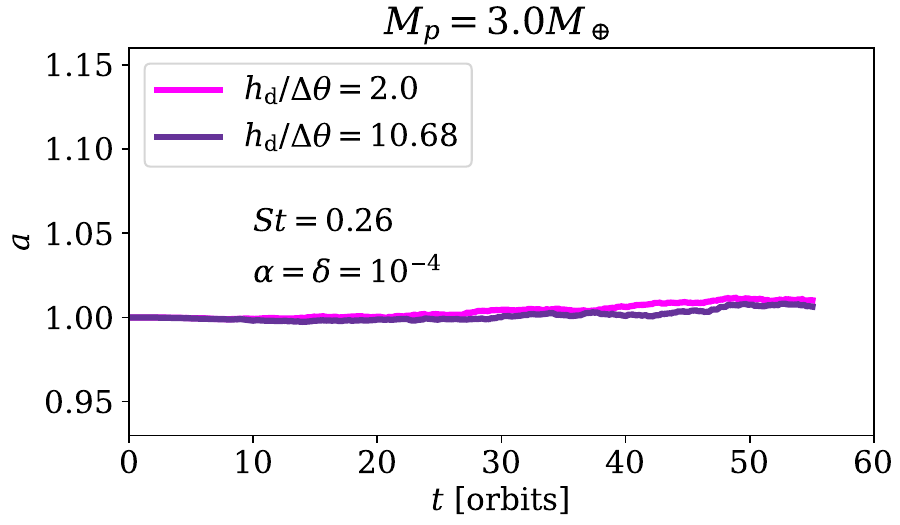}
    \caption{Comparison of the temporal evolution of the semimajor axis of a planet of mass $M_p=3.0M_\oplus$ for $\alpha=\delta=10^{-4}$ and $\mathrm{St}=0.26$ using prescriptions 1) and 2) for the resolution of the grid in the vertical component (magenta and purple lines, respectively).}
    \label{fig:comp_res}
\end{figure}

Note that, due to the high computational cost of our 3D models when increasing the number of zones in the vertical component of the grid, we have fixed the $h_\mathrm{d}/\Delta\theta$ ratio to a little more than at least ten grid cells for $h_\mathrm{d}$. Nevertheless, we think that it is a good compromise to keep this minimum value for the numerical resolution of $h_\mathrm{d}$. In Fig. \ref{fig:comp_res} we show the comparison of the temporal evolution of the semi-major axis of a planet with $M_p=3M_\oplus$ for the pair of parameters $(\mathrm{St}, h_\mathrm{d})=(0.26,10^{-4})$, applying both grid configurations. It is clear that the change is minimal between the two prescriptions of the vertical grid. Therefore, it is safe to adopt this vertical resolution ratio in description 2).

\end{appendix}

%
\bibliographystyle{aa} 
\bibliography{manuscript.bib} 
%



\end{document}